\documentclass[letter]{aa}  

\usepackage{natbib}
\usepackage{graphicx}
\usepackage{txfonts}
\usepackage{hyperref}
\begin{document} 

\title{Constraints on the cosmological coupling of black holes from Gaia}

\author{
    Ren\'e Andrae\orcit{0000-0001-8006-6365}\inst{\ref{inst:0001}}\thanks{andrae@mpia-hd.mpg.de}
    \and
    Kareem El-Badry\inst{\ref{inst:0002},\ref{inst:0001}}
}

\institute{
Max Planck Institute for Astronomy, K\"{ o}nigstuhl 17, 69117 Heidelberg, Germany\relax\label{inst:0001}
\and
Center for Astrophysics | Harvard \& Smithsonian, 60 Garden Street, Cambridge, MA 02138, USA\relax\label{inst:0002}
}

\date{Submitted 8th March 2023; accepted 2nd May 2023}

 
\abstract{
Recent work has suggested that black holes (BHs) could be cosmologically coupled to the accelerated expansion of the universe, potentially becoming a candidate for dark energy. This would imply BH mass growth following the cosmological expansion, with the masses of individual BHs growing as $M_{\rm BH}\propto (1+z)^3$. In this letter, we discuss the binary systems Gaia BH1 and Gaia BH2, which contain $\sim 9\,M_{\odot}$ BHs orbited by $\sim 1\,M_{\odot}$ stars in widely-separated orbits. The ages of both systems can be constrained by the properties of the luminous stars.  If BH masses are indeed growing as $(1+z)^3$, the masses of both BHs at formation would have been significantly smaller than today. We find a 77\% probability that the mass of the BH in Gaia BH2 would have been below $2.2M_\odot$ at formation. This is below the classical Tolman-Oppenheimer-Volkov limit, though it is not yet clear if BHs subject to cosmological coupling should obey this limit. For Gaia BH1, the same probability is 70\%. This analysis is consistent with results from two BHs in the globular cluster NGC~3201, but unlike the NGC~3201 BHs, the Gaia BHs have well-constrained inclinations and thus firm upper mass limits. The discovery of more BHs in binary systems with {\it Gaia} astrometry in the coming years will allow us to test the cosmological coupling hypothesis decisively.
}


\providecommand{\orcit}[1]{\protect\href{https://orcid.org/#1}{\protect\includegraphics[width=8pt]{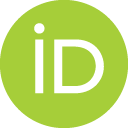}}}

\maketitle
%

\section{Introduction}
\label{sect:introduction}

Black holes (BHs) are objects of great interest in astrophysics as final stages of stellar evolution \citep[e.g.][]{2018ApJ...852L..19C} as well as active Galactic nuclei being visible already in the earliest times of the Universe \citep[e.g.][]{Wang_2021}. BHs are often described within the Schwarzschild or Kerr solutions of general relativity. However, those solutions approach a flat spacetime at infinite spatial distance from the BH. This is inconsistent with the standard cosmological assumption that on large scales spacetime approaches a Friedman-Robertson-Walker (FRW) metric. Since the field equations of general relativity are nonlinear, this cannot be resolved with a simple superposition of a ``local'' Kerr solution and a ``global'' FRW metric.

Finding BH solutions that approach an FRW metric at infinte spatial distance is the subject of ongoing research. In such a solution, the relativistic material (i.e.\ the BH) can become coupled to the cosmological expansion of the Universe \citep[e.g.][]{2007PhRvD..76f3510F,2019ApJ...882...19C}. This would affect BHs that contain vacuum energy, which may apply to stellar BHs as well as super-massive BHs (SMBHs).

Recently, \citet{2023ApJ...943..133F} report that they find SMBHs in quiescent elliptical galaxies to grow in mass by factors 7-20 between cosmological redshifts $0.7\lesssim z\lesssim 2.5$ and $z\sim 0$, while at the same time no growth in stellar mass is observed.  \citet{2023ApJ...944L..31F} seek to explain this observation with cosmological coupling of the SMBH masses of the form
\begin{equation}\label{eq:coupled-mass-with-a}
M(a) = M(a_i) \left(\frac{a}{a_i}\right)^k ,
\end{equation}
where $M$ denotes the mass of the BH, $a$ and $a_i$ denote the scale factors at the current time and the initial time when the BH formed, and $k$ is the cosmological coupling strength. Using $a=1/(1+z)$ and assuming current time $z=0$ for observation, we can translate Eq.~(\ref{eq:coupled-mass-with-a}) into cosmological redshift such that
\begin{equation}\label{eq:coupled-mass}
M(z=0)=M(z_i)\,(1+z_i)^k,
\end{equation}
with $z_i$ denoting the redshift when the BH formed. Within the physically realistic range of $-3\leq k\leq +3$ \citep[e.g.][]{2021ApJ...921L..22C}, different values of $k$ are plausible: $k=0$ corresponds to the scenario where there is no coupling at all. An analysis of the LIGO/Virgo BH mergers by \citet{2021ApJ...921L..22C} estimates $k\sim 0.5$, which is also compatible with results from \citet{2023arXiv230212386R} assuming that there is no mass gap between neutron stars and BHs. Then there is the comoving scenario of $k=1$ where the Schwarzschild radius expands together with the scale factor \citep{2007PhRvD..76f3510F}. Finally, $k=3$ is the scenario producing a constant BH energy density. Given the results in \citet{2023ApJ...943..133F}, \citet{2023ApJ...944L..31F} estimate $k=3.11_{-1.33}^{+1.19}$, and they emphasise that they exclude the no-coupling scenario of $k=0$ at 99.98\% confidence ($\sim 3.9\sigma$), including random errors as well as their best estimate of systematic effects. In particular, \citet{2023ApJ...944L..31F} suggest that also stellar-mass BHs are subject to such cosmological coupling and may be related (potentially even causally) to the accelerated expansion of the Universe. They also propose that cosmologically-growing stellar BHs formed at early times could be the progenitors of SMBHs observed later.

Taking the results reported by \citet{2023ApJ...943..133F} at face value, we want to investigate if a cosmological coupling of BH masses of the form of Eq.~(\ref{eq:coupled-mass}) is consistent with observed stellar-mass BHs in binary systems. It is generally assumed that in such binaries the BH progenitor and the star that survives today formed at (nearly) the same point in time from the same gas reservoir. The more massive primary star quickly evolves into a BH within at most a few tens of million years. The less massive secondary, on the other hand, evolves much more slowly and is still observable today. We are particularly interested in binary systems which we can date, either through the visible secondary star or through the binary system being part of a larger stellar system of known age. This allows us to infer the masses of the BHs at the time of their formation, under the assumption of $k=3$ cosmological coupling. Stellar-remnant BHs are not expected to form with masses below the Tolman-Oppenheimer-Volkoff (TOV) limit of $\approx 2.2\,M_{\odot}$ \citep[e.g.][]{Rezzolla2018} because the would-be progenitors of such BHs become neutron stars or white dwarfs instead. However, given the unknowns in the equation of state for neutron stars, the exact value of the TOV limit remains somewhat uncertain: \citet{2021arXiv210708822R} suggest that it could be as high as $2.6\,M_{\odot}$ and it is conceivable to have even more massive neutron stars supported by fast rotation \citep{2018Sci...359..724C,Rezzolla2018}. Inferred initial BH masses below the TOV limit would thus disfavor the cosmological coupling hypothesis.
Additionally, there may be a mass gap between the most massive NS and the least massive BH \citep[e.g.][]{2010ApJ...725.1918O,2011ApJ...741..103F}, i.e.\ between the TOV limit and a possible minimal BH mass of $5.4\,M_\odot$ \citep{2022ApJ...937...73Y}. However, the existence of such a mass gap is under dispute.

In this letter, we are investigating the binary system Gaia~BH2, for which astrometry and spectroscopy allowed \citet{2023arXiv230207880E} to directly infer the mass of the BH (without any ambiguity due to unknown inclination angle). In Sect.~\ref{sect:GaiaBH2}, we estimate the age of Gaia~BH2, which allows us to infer the original BH mass under the cosmological-coupling hypothesis. In Sect.~\ref{sect:discussion}, we compare our findings to other results and briefly discuss further possible tests of the cosmological coupling hypothesis.

\section{Gaia BH2}
\label{sect:GaiaBH2}

Using Gaia DR3 data \citep{2022arXiv220800211G} and follow-up spectroscopy, \citet{2023arXiv230207880E} report finding a binary system comprised of a red-giant star and a BH with mass $M_{\rm BH}=(8.9\pm 0.3)M_\odot$. The red-giant star has an apparent magnitude of $G\sim 12.3$~mag and resides slightly above the Milky Way disk ($b\sim 2.8^\circ$) and about $50^\circ$ away from the Galactic center. Despite its low Galactic latitude, \citet{2023arXiv230207880E} estimate the red-giant star's reddening as only $E(B-V)\sim 0.2$~mag, which is most likely due to it being only 1.16~kpc away.

\citet{2023arXiv230207880E} consider it unlikely that the red giant has undergone significant mass transfer to the BH, given the binary’s wide and non-circular orbit, the lack of evidence for ongoing mass transfer (no counterpart was found in X-ray or radio), and the prediction that any mass transfer from the BH progenitor to the secondary would have been unstable and short-lived. This also makes it very unlikely that the black hole started out as a neutron star and formed only later through mass transfer.

We refine the original analysis of the secondary star in \citet{2023arXiv230207880E} by re-fitting its broadband SED using an MCMC algorithm. We place a Gaussian prior on the metallicity motivated by their spectroscopic observations, with $\rm [M/H]\sim \mathcal{N}(-0.02, 0.05)$, and a prior on the extinction of $E(B-V)\sim \mathcal{N}(0.2, 0.03)$. We also leave the distance as a free parameter, constrained by the {\it Gaia} parallax. Our approach is quite similar to that adopted by \citet{2023arXiv230207880E}, but with the advantage that we track correlations between parameters, most significantly, the star's temperature and radius. This produces the estimates $T_\textrm{eff}=4627\,K$,  $R=7.73\,R_\odot$ and $\textrm{[M/H]}=-0.021$ and the following covariance matrix (in the same order of parameters):
\begin{equation}
\Sigma = \left(\begin{array}{ccc}
4957 & -8.26 & -0.076 \\
-8.26 & 0.061 & 0.00057 \\
-0.076 & 0.00057 & 0.00094
\end{array}\right),
\end{equation}
where the numbers on the diagonals correspond to the squared uncertainties of each parameter. Our constrains are consistent within 1$\sigma$ with those reported by \citet{2023arXiv230207880E}.

\begin{figure}
\begin{center}
\includegraphics[width=\columnwidth]{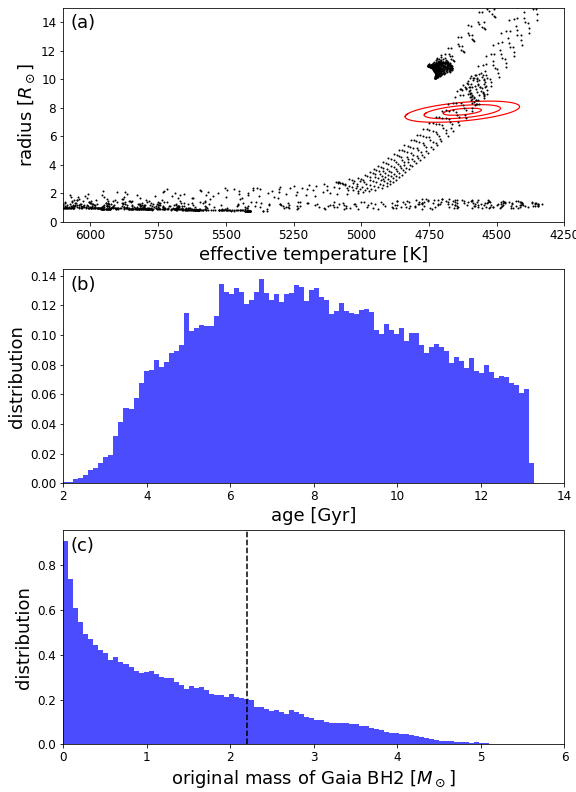}
\end{center}
\caption{Age determination of Gaia~BH2. Panel a: temperature-radius diagram with Gaia~BH2 marked by red error ellipses of 1, 2, $3\sigma$, PARSEC isochrones for $\textrm{[M/H]}=-0.0244$ in black. Panel b: Age distribution estimated from Monte-Carlo sampling. Panel c: Distribution of original mass of Gaia~BH2, which is current mass divided by $(1+z)^3$ according to Eq.~(\ref{eq:coupled-mass}). The vertical dashed line indicates $2.2M_\odot$, below which a BH is unlikely to form.}
\label{fig:RGB-star-vs-isochrone}
\end{figure}

Using isochrones from PARSEC 1.2S Colibri S37 models \citep[][and references therein]{2014MNRAS.445.4287T,2015MNRAS.452.1068C,2020MNRAS.498.3283P} and the isochrone forward model described in \citet{2022arXiv220606138A}, we draw Monte-Carlo samples from the likelihood specified by the temperature, radius and [M/H] estimates and their covariance above, where the age is one of the free model parameters. An example isochrone is shown in Fig.~\ref{fig:RGB-star-vs-isochrone}a and Fig.~\ref{fig:RGB-star-vs-isochrone}b shows the resulting age distribution for the red giant in Gaia~BH2. Its median age estimate is 7.9~Gyr (5.2 to 11.1~Gyr central 68.3\% confidence interval, 3.5 to 12.8~Gyr central 95.5\% confidence, 2.5 to 13.5~Gyr central 99.7\% confidence). This is consistent with the $\alpha$-high chemical composition found spectroscopically by \citet{2023arXiv230207880E}. Neglecting the time it took the massive primary star to evolve into a BH, we can approximate the BH formation with the age of the red giant star. 

Using the results from \citet{2020A&A...641A...6P}, we can convert the age distribution from Fig.~\ref{fig:RGB-star-vs-isochrone}b into a distribution of cosmological redshift $z_{\rm form}$ at which the formation of the BH would have occurred and from that we obtain the distribution of $(1+z_{\rm form})^3$. Quantitatively, we find a median mass inflation factor of $(1+z_{\rm form})^3\sim 8.0$ from our Monte-Carlo samples, with a central 68.3\% confidence interval ranging from 3.4 to 39.8 (2.2 to 321 central 95.5\% confidence, 1.7 to 765 central 99.7\% confidence). We finally obtain the distribution of Gaia~BH2's original mass shown in Fig.~\ref{fig:RGB-star-vs-isochrone}c by dividing the current mass by $(1+z_{\rm form})^3$. Here, we also propagate the uncertainty of the current mass of Gaia~BH2, $(8.9\pm 0.3)M_\odot$, by drawing Monte-Carlo samples. As is evident from Fig.~\ref{fig:RGB-star-vs-isochrone}c, most of these original masses (76.92\%) are lower than $2.2\,M_\odot$.
%
%
Specifically, the median formation mass estimate is 1.11$M_\odot$ with a central 68.3\% confidence interval ranging from 0.223 to 2.64$M_\odot$ (0.028 to 4.07$M_\odot$ central 95.5\% confidence, 0.012 to 5.17$M_\odot$ central 99.7\% confidence). This suggests that --  if cosmological coupling with $k=3$ applied -- the original mass at the time of BH formation would have been lower than the mass of any known BH, and most likely, too low to actually form a BH through known astrophysical channels.
Table~\ref{table:my_label} provides the confidence levels rejecting various other values of $k$.

\begin{table}[]
\begin{center}
\caption{Fraction of Monte-Carlo samples for which the original BH mass at formation is below $2.2\,M_\odot$ for various coupling constants $k$ (see Sect.~\ref{sect:introduction} for explanations of these different $k$ values).}
\label{table:my_label}
\begin{tabular}{c|c|c|c|c}
$k$  &  0  &  0.5  &  1  &  3 \\
\hline
Gaia BH1  &  0.0  &  0.0  &  0.0126  &  0.6996 \\
Gaia BH2  &  0.0  &  0.0  &  0.1086  &  0.7692 \\
\end{tabular}
\end{center}
\end{table}

\section{Gaia BH1}
\label{sect:GaiaBH1}

We repeat the exact same procedure for Gaia~BH1 \citep{GaiaBH1}, which is a binary system of a black hole of mass $(9.62\pm 0.18)M_\odot$ and a main-sequence G-dwarf that is slightly hotter than the Sun and has a lower metallicity of $\textrm{[M/H]}=-0.2$. Its apparent magnitude is $G\sim 13.8$~mag and it is 480~pc away. With Galactic coordinates $(\ell,b)=(22.63^\circ, 18.05^\circ)$ it is also well above the Galactic plane. \citet{GaiaBH1} estimate its reddening as $E(B-V)\sim 0.29$.

\begin{figure}
\begin{center}
\includegraphics[width=\columnwidth]{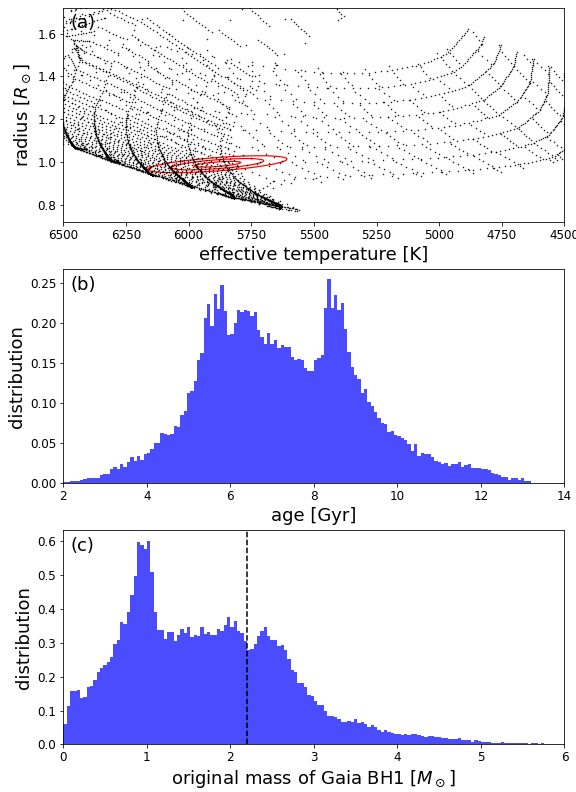}
\end{center}
\caption{Same as Fig.~\ref{fig:RGB-star-vs-isochrone} but for Gaia~BH1. The PARSEC isochrone in panel (a) if for $\textrm{[M/H]}=-0.1963$.}
\label{fig:RGB-star-vs-isochrone-Gaia-BH1}
\end{figure}

Re-analysing Gaia~BH1 using the same approach as for Gaia~BH2, we estimate $T_\textrm{eff}=5885\,K$, $R=0.99\,R_\odot$ and $\textrm{[M/H]}=-0.1963$ with the following covariance matrix:
\begin{equation}
\Sigma = 
\left(\begin{array}{ccc}
8547 & -0.742 &  -0.358 \\
-0.742 &  0.000173 & 0.0000601 \\
-0.358 & 0.0000601 & 0.000515 
\end{array}\right)   
\end{equation}
Again, using the isochrone forward model from \citet{2022arXiv220606138A}, we draw Monte-Carlo samples from the likelihood in $(T_\textrm{eff}, R, \textrm{[M/H]})$ space and thus infer the age of Gaia~BH1, which we then convert to redshift using \citet{2020A&A...641A...6P}. The results are shown in Fig.~\ref{fig:RGB-star-vs-isochrone-Gaia-BH1}. Despite the visible secondary being a G~dwarf on the main sequence, we find the age to be reasonably well constrained with a median age of 7.1~Gyr (5.4 to 9.1~Gyr central 68.3\% confidence interval, 3.8 to 11.4~Gyr central 95.5\% confidence, 2.3 to 12.8~Gyr central 99.7\% confidence). Thus, our age estimate of Gaia~BH1 is younger and more precise than Gaia~BH2. The reason for this surprisingly precise age estimate is that the secondary G~dwarf is not exactly on the main sequence but rather appears to be entering the turn-off phase, as is evident from Fig.~\ref{fig:RGB-star-vs-isochrone-Gaia-BH1}a. Here, we emphasise that the uncertainty of the age estimate is mainly propagated from the atmospheric parameters, i.e.\ the uncertainty of the secondary genuinely being in the turn-off phase vs.\ it still being on the main sequence is included. Concerning the original BH mass, Fig.~\ref{fig:RGB-star-vs-isochrone-Gaia-BH1}c, it was below $2.2\,M_\odot$ with 69.96\% confidence.
%
%
As with Gaia~BH2, this suggests that --  if cosmological coupling with $k=3$ applied -- the original mass at the time when Gaia~BH1 formed would have been lower than the mass of any known BH, and most likely, too low to actually form a BH through known astrophysical channels.
Again, Table~\ref{table:my_label} provides the confidence levels rejecting various other values of $k$.

\section{Discussion}
\label{sect:discussion}

\begin{figure}
\begin{center}
\includegraphics[width=\columnwidth]{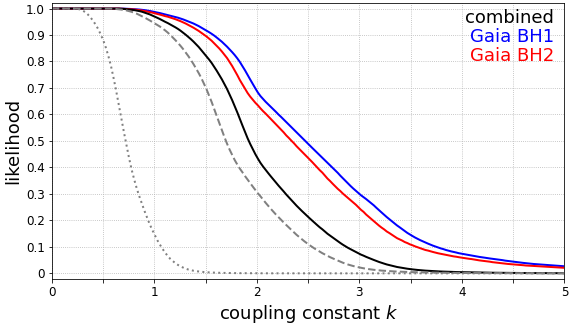}
\end{center}
\caption{Likelihood of original BH mass at formation being below the TOV limit of $2.2\,M_\odot$ as function of coupling constant $k$. Blue line is for Gaia~BH1, red for Gaia~BH2 and black for the combination of both. Since the measurements of Gaia~BH1 and BH2 are statistically independent, the combination is simply the product of the two individual likelihoods. Additionally, the grey dashed line shows the combined results for a higher TOV limit of $2.6\,M_\odot$ \citep{2021arXiv210708822R} while the grey dotted line shows the combined results for a mass gap with a minimal BH mass of $5.4\,M_\odot$ \citep{2022ApJ...937...73Y}.}
\label{fig:estimation-of-k}
\end{figure}

If the cosmological coupling strength of $k=3$ applied, it would imply an original BH mass below the TOV limit of $2.2M_\odot$ at 76.92\% confidence for Gaia~BH2 and 69.96\% confidence for Gaia~BH1. Since both estimates for Gaia~BH1 and BH2 are statistically independent, the combined probability that both BHs could have been above the TOV limit is only 6.9\%. Stellar cores below $\sim 1.4M_\odot$ will rather form a white dwarf and cores below $\sim 2.2M_\odot$ will rather form a neutron star, but not a BH. In fact, \citet{2021arXiv210708822R} suggest that neutron stars could remain stable up to $2.5-2.6M_\odot$. Consequently, Gaia~BH1 and BH2 disfavour the cosmological coupling with $k=3$ put forward by \citet{2023ApJ...944L..31F}.
Table~\ref{table:my_label} provides the probabilities for additional values of coupling constants $k$ and Fig.~\ref{fig:estimation-of-k} shows a graphical representation of how Gaia~BH1 and BH2 constrain $k$. While Fig.~\ref{fig:estimation-of-k} cannot distinguish between values of $k\lesssim 1$, it clearly disfavours $k>3$, which is outside of the physically realistic range \citep[e.g.][]{2021ApJ...921L..22C}. If we allow for a higher TOV limit of $2.6\,M_\odot$ or a mass gap between NS and BHs reaching to $5.4\,M_\odot$, the likelihood of $k>1$ decreases even further.
Our finding is in agreement with similar results reported by \citet{2023arXiv230212386R}, who has used two dynamically confirmed BHs in the globular cluster NGC~3201 to show that both of these BHs would need to have near face-on inclinations in order to be consistent with the $k=3$ cosmological coupling, which is a configuration that \citet{2023arXiv230212386R} assessed to have a probability of at most $10^{-4}$. 
While the results from \citet{2023arXiv230212386R} are considerably more restrictive, nearby astrometrically-characterized BHs like Gaia~BH1 and BH2 offer constraints on BH coupling that are complimentary to systems in globular clusters: their masses are better-constrained, but their ages are more uncertain. Furthermore, globular clusters often contain blue stragglers formed by stellar encounters. While we are not suggesting that the two BHs in NGC~3201 have necessarily formed this way, it is in principle possible that BHs in globular clusters form later through collisions or capture of a neutron star followed by accretion. This is extremely unlikely for BHs detected astrometrically in the field outside of globular clusters.

At the fundamental level, \citet{2023arXiv230212386R} and our work rely heavily on the assumption that cosmological coupling and the TOV limit both apply at the same time to stellar-remnant black holes. This may not be the case as the TOV limit theoretically requires the existence of an event horizon, which may not be the case for singularity-free BHs containing vacuum energy. However, as discussed in \citet{2023arXiv230212386R} (Sect.~4.1 therein), the formation of horizonless BHs of masses below the TOV limit is difficult to reconcile with observations of BHs in high-mass X-ray binaries.

A stronger conclusion from Gaia~BH1 and Gaia~BH2 would require a more accurate age estimate of the RGB companion in Gaia~BH2. Alternatively, the discovery of more BHs in binary systems using Gaia astrometry is likely to provide additional constraints. \citet{2023MNRAS.518.1057E} estimate that Gaia~DR4 will allow to astrometrically find a few dozens of BHs in binary systems.

We finally note that we know of young BHs with masses $M>10M_\odot$ in high-mass X-ray binaries like Cygnus X-1 \citep[e.g.][]{Miller-Jones2021}. It is reasonable to assume that such massive stellar BHs should also have formed in the early Universe. If a $k=3$ coupling applied, we should eventually find BHs in the mass range $70-160M_\odot$, where our current understanding of pair-instability supernovae predicts that there actually should be no BHs formed by stellar evolution \citep{2021ApJ...912L..31W}. Relatedly, {\it if} $k=3$ cosmological coupling holds and Gaia BH2 formed with a mass of e.g. $3\,M_{\odot}$, then {\it Gaia} should soon discover many BHs with similar masses that formed more recently.
It is thus likely that {\it Gaia} will be able to test the cosmological coupling of BHs and their potential link to Dark Energy conclusively within the next few years.

\begin{acknowledgements}
We thank the anonymous referee for very useful feedback that greatly improved this manuscript. RA also thanks Knud Jahnke for detailed discussions on selection biases and their possible impact on SMBH masses in ellipticals at different redshifts.
\end{acknowledgements}

\bibliographystyle{aa} 
\bibliography{andrae-elbadry-2023} 

\end{document}